\documentclass[12pt]{iopart}
\usepackage{iopams}
\usepackage{enumerate}
\usepackage{amsfonts}
\usepackage{amssymb}
\usepackage{amsthm}
\usepackage{braket}
\usepackage{mathrsfs}
\usepackage{dcolumn}
\usepackage{bm}
\usepackage{graphicx, graphics}
\usepackage{color}
\usepackage[pdftex]{hyperref}
\hypersetup{
colorlinks=true,
linkcolor=[rgb]{0.0,0.1,0.8},
citecolor=blue,
filecolor=magenta,
urlcolor=[rgb]{0.0,0.3,1}
}

 \newcommand{\pap}[1]{\left( #1 \right)}

\begin{document}
\title[Squeezing generation crossing a mean-field critical point]{
Squeezing generation crossing a mean-field critical point: Work statistics, irreversibility and critical fingerprints}

\author{Fernando J. G{\'o}mez-Ruiz$^{1,2,\dagger}$\href{https://orcid.org/0000-0002-1855-0671}{\includegraphics[scale=0.5]{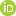}},
Stefano Gherardini$^{3,4}$\href{https://orcid.org/0000-0002-9254-507X}{\includegraphics[scale=0.5]{orcid}},
Ricardo Puebla$^{1,\star}$\href{https://orcid.org/0000-0002-1243-0839}{\includegraphics[scale=0.5]{orcid}}
}
\address{$^1$ Departamento de F{\'i}sica, Universidad Carlos III de Madrid, Avda. de la Universidad 30, 28911 Legan{\'e}s, Spain}
\address{$^2$ Departamento de F\'isica Te\'orica, At\'omica y \'Optica and  Laboratory for Disruptive Interdisciplinary Science, Universidad de Valladolid, 47011 Valladolid, Spain}
\address{$^3$ Istituto Nazionale di Ottica del Consiglio Nazionale delle Ricerche (CNR-INO), Largo Enrico Fermi 6, I-50125 Firenze, Italy}
\address{$^4$ European Laboratory for Non-linear Spectroscopy, Universit{\`a} di Firenze, I-50019 Sesto Fiorentino, Italy}
\ead{$\dagger$ \href{mailto:fegomezr@fis.uc3m.es}{fegomezr@fis.uc3m.es}\\
$\star$ \href{mailto:rpuebla@fis.uc3m.es}{rpuebla@fis.uc3m.es}}

\begin{abstract}
Understanding the dynamical consequences of quantum phase transitions on thermodynamical quantities, such as work statistics and entropy production, is one of the most intriguing aspect of quantum many-body systems, pinpointing the emergence of irreversibility to critical features. In this work, we investigate the critical fingerprints appearing in these key thermodynamical quantities for a mean-field critical system undergoing a finite-time cycle, starting from a thermal state at a generic inverse temperature. In contrast to non-zero dimensional many-body systems, the presence of a mean-field critical point in a finite-time cycle leads to constant irreversible work even in the limit of infinitely slow driving. 
This links with the fact that a slow finite-time cycle results in a constant amount of squeezing, which enables us to derive analytical expressions for the work statistics and irreversible entropy, depending solely on the mean-field critical exponents and the functional form of the control parameter near the critical point. 
We find that the probability of observing negative work values, corresponding to negative irreversible entropy, is inversely proportional to the time the system remains near to the critical point, and this trend becomes less pronounced the lower the temperature of the initial thermal state. 
Finally, we determine the irreversibility traits under squeezing generation at zero-temperature using the relative entropy of coherence.
\end{abstract}
\\
\submitto{\href{https://iopscience.iop.org/article/10.1088/2058-9565/adf5de}{Quantum Sci. Technol. {\bf 10}, 045011 (2025)}}
\maketitle

\section{Introduction}\label{s:intro}

The intersection of quantum information and quantum thermodynamics has found profound implications, positioning information as a resource comparable to traditional thermodynamic resources such as heat and work~\cite{Deffner,Landi:21}. The primary goal of quantum thermodynamics consists in understanding how standard thermodynamic quantities are altered when quantum effects become significant~\cite{Binder,Oppenheim:02,GherardiniTutorial}. These quantities include entropy production~\cite{Andolina:19,Andolina:19b,CimininpjQI:20,Belenchia:20,Ptaszynski:23,Hernandez-Gomez:23}, work statistics~\cite{Dorner:12, Funo:17, Zawadzki:20, hernandezProj:24, hernandezInterfero:24}, and the laws of thermodynamics~\cite{Brandao:08, Buffoni:22}. This burgeoning field has garnered substantial interest as researchers strive to unravel the complex behaviors that emerge in the quantum realm, where traditional classical descriptions fall short~\cite{Vicente:24}. Phenomena such as quantum fluctuations, coherence, and entanglement~\cite{Horodecki:09, Amico:08} necessitate new theoretical frameworks to adequately describe these systems.

Quantum phase transitions (QPTs) are foundational to condensed matter physics and quantum technologies, presenting some of the most intriguing aspects of quantum many-body systems~\cite{Polkovnikov:11}. Unlike classical phase transitions driven by thermal fluctuations at finite temperatures, QPTs occur at absolute zero, driven by quantum fluctuations that can destroy long-range order~\cite{Sachdev}. The nonequilibrium dynamics triggered by QPTs have attracted significant attention, especially in light of recent experimental advances in ultracold atomic and molecular gases, trapped ions, and solid-state systems. These advances enable high degrees of isolation and control, opening new avenues for exploiting quantum effects in various applications, including quantum computation~\cite{Preskill:18}, information processing, quantum sensing~\cite{Degen:17}, and quantum thermodynamics~\cite{Landi:21}.

One particularly exciting area of research involves the finite-time modulation of control parameters in quantum systems~\cite{Talkner:08, Dutta:15, Perfetto:19, Yadalam:19, Heerwagen:20, Yi:11, Deffner:15}, which is essential for applications such as quantum thermodynamic engines~\cite{Watanabe:17, Deng:18, Watanabe:20}, quantum annealing~\cite{Bando:20, King:22}, and adiabatic quantum computation. However, finite-time evolutions often lead to nonadiabatic excitations, posing performance challenges. This issue is exacerbated when the evolution passes through or reaches a QPT, where the vanishing energy gap at the critical point disrupts the adiabatic condition, causing excitations~\cite{Kibble:76, Kibble:80, Zurek:85, Polkovnikov:08, delCampo:10, Puebla:19, GomezRuiz:20, Gherardini:24, Puebla:20}. The finite-time evolution across a QPT is responsible for significant changes in the statistical properties of energy-change distribution as its skewness~\cite{Zawadzki:20}.

Despite these challenges, quantum many-body systems hold great promise as the working medium for heat engines and quantum batteries~\cite{Campaioli:24}, offering potential enhancements in thermodynamic performance through quantum effects~\cite{Varizi:20, Francica:19, Abah:22}. Understanding the scenarios in which QPTs can be advantageous is crucial for developing strategies to scale up quantum machines~\cite{Bando:20} and achieve quantum enhancements in various thermodynamic tasks.

In this work, we delve into the critical fingerprints left on key thermodynamic quantities by mean-field critical systems undergoing finite-time cycles. These systems have been less explored compared to their non-zero dimensional many-body counterparts, where the inter-particle interactions depend on the distance among them, such as in the paradigmatic Ising model with nearest-neighbors couplings. In contrast to these non-zero dimensional systems, we demonstrate that mean-field critical points in finite-time cycles lead to a constant irreversible work even with infinitely slow driving. Albeit this finding has been recently discussed in \cite{Gherardini:24} for cycles initialized in the vacuum state, we extend the analysis to thermodynamic cycles starting from a thermal state at generic inverse temperature $\beta$. Moreover, we show that the irreversibility of a mean-field system proportionally increases with the time it stays close to the critical point, and this trend grows with increasing $\beta$. In addition, our results include analytical expressions for work statistics and irreversible entropy that are directly derived from mean-field critical exponents and the shape of the control parameter near the critical point. Finally, we analyze how the quantum coherence originated by the thermodynamic cycle affects its irreversibility while crossing the mean-field critical point.

The article is organized as follows. First, in Sec.~\ref{s:model} we introduce the mean-field critical model and the details regarding the time-dependent protocol to realize a cycle. In Sec.~\ref{s:Sirr} we analyze the universal behavior of the irreversible work and entropy, which are completely determined by both the critical exponents and the form of the protocol in the vicinity of the critical point. In Sec.~\ref{s:coh} we study the role that quantum coherence plays in irreversible entropy, while in Sec.~\ref{s:conc} we summarize the main findings of the article.

\section{Model and protocol}\label{s:model}

We shall start our analysis considering a driven quantum harmonic oscillator ($\hbar=1$) with Hamiltonian
\begin{equation}\label{eq:H0}
     \hat{H}(t)=\omega \, \hat{a}^\dagger \hat{a} - \omega\frac{g^2(t)}{4}\pap{\hat{a}+\hat{a}^\dagger}^2,
\end{equation}
which is described in terms of the standard creation and annihilation operators, $[\hat{a},\hat{a}^\dagger]=1$, and is driven according to a cyclic transformation reaching the final value $g_f$. The Hamiltonian of Eq.~(\ref{eq:H0}) effectively captures the critical features of different fully-connected models such the Lipkin-Meshkov-Glick model~\cite{Lipkin:65, Ribeiro:07, Ribeiro:08}, the critical quantum Rabi model and related systems displaying a superradiant phase transition~\cite{Dicke:54, Emary:03, Emary:03prl, Hwang:15, Puebla:16, Peng:19, Zhu:20}, as well as different realizations of Bose-Einstein condensates~\cite{Anquez:16, Mottl:12, Brennecke:13, Zibold:10}. It is worth mentioning that these models have been realized experimentally~\cite{Anquez:16, Mottl:12, Brennecke:13, Zibold:10, Jurcevic:17, Cai:21}. Our goal consists in exploring the impact of a mean-field QPT in relevant thermodynamic quantities and how irreversibility emerges in the exact thermodynamic limit of these models. This can be studied by considering the Hamiltonian $\hat{H}(t)$ in Eq.~(\ref{eq:H0}), which represents a valid description of these models in this limit for $|g|\leq g_c=1$ [see for example Refs.~\cite{Hwang:15,Puebla:20,Abah:22,Gherardini:24} for the details on how a quantum Rabi and a Lipkin-Meshkov-Glick model reduce to Eq.~(\ref{eq:H0})] where $g_c$ denotes the critical point at which the energy gap $\epsilon(g) =\omega\sqrt{1-g^2}$ vanishes as $\epsilon(g)\approx |g-g_c|^{z\nu}$ with $z\nu=1/2$  the critical exponents of the mean-field QPT~\cite{Sachdev}.

In the following, we consider that the time-dependent protocol reads as
\begin{equation}\label{eq:gt}
 g(t)=\Bigg\lbrace
 \begin{array}{cc}
 g_f\left[1-\pap{\frac{\tau-t}{\tau}}^r\right]&{\rm for}\; 0\leq t \leq \tau,\\[1ex]
 g_f\left[1-\pap{\frac{t-\tau}{\tau}}^r \right]&{\rm for}\; \tau < t \leq 2\tau,
 \end{array}
\end{equation}
where we have assumed, without loss of generality, that $g(0)=g(2\tau)=0$. The nonlinear exponent $r>0$ controls how the system approaches the final coupling $g_f$. For $r=1$ the cycle is performed linearly. For $g_f=g_c$, the different exponents $r$ modifies the Kibble-Zurek scaling laws~\cite{Barankov:08} since $|g_c-g(t)|\propto |t-\tau|^r$, so that the rate at which the system is driven is modified as $|\dot{g}(t)|=2g_c r|t-\tau|^{r-1}\tau^{-r}$. Note that the larger the nonlinear exponent $r$, the longer the system stays near the critical point. We would also stress that the dynamics originated by the time-dependent protocol for $\tau < t \leq 2\tau$ has not to be considered as the time-reversal of the time-evolution due to applying $g(t)$ for $0\leq t \leq \tau$.

As reported in Refs.~\cite{Defenu:21,Gherardini:24}, for a cycle that reaches the critical point $g_f=g_c$ with duration such that $2\tau\omega\gg 1$, the initial state is never retrieved regardless of how slowly the protocol is executed. In particular, we consider as an initial state the thermal state, 
\begin{equation}
 \hat{\rho}(0)=\hat{\rho}_\beta=\frac{ \exp[-\beta \omega \hat{a}^\dagger \hat{a}] }{ Z_\beta }
\end{equation}
with $Z_\beta={\rm Tr}[\exp[-\beta \omega \hat{a}^\dagger \hat{a}]]$ being $\beta$ the inverse temperature. This thermal state contains $N_\beta\equiv {\rm Tr}[\hat{\rho}_\beta\hat{a}^\dagger \hat{a}]=(e^{\beta\omega}-1)^{-1}$ thermal excitations. In the low-temperature regime, $\beta\omega\gg 1$, the $N_\beta$ decreases exponentially in $\beta\omega$, $N_\beta\sim e^{-\beta\omega}$, while in the high-temperature limit, $\beta\omega\ll 1$, the number of thermal excitations grows as $N_\beta\sim (\beta\omega)^{-1}$. These initial thermal states become squeezed upon the completion of the cycle at time $2\tau$, with a squeezing amplitude $|s|$ only determined by the critical exponents $z\nu$ and $r$~\cite{Abah:22, Garbe:22}:
\begin{equation}\label{eq:s}
      |s| = {\rm arcosh}\left[\csc\left[\frac{\pi}{2(1+z\nu r)}\right]\right],
\end{equation}
where we explicitly left the critical exponents $z\nu$ to remark how different quantities inherit its dependence, although $z\nu=1/2$ for the mean-field QPT. At the completion of the cycle, $\hat{\rho}(2\tau)=\hat{\mathcal{S}}(s)\hat{\rho}_\beta \hat{\mathcal{S}}^\dagger(s)$ where $\hat{\mathcal{S}}(s)=\exp[(s^*\hat{a}^2-s (\hat{a}^\dagger)^2)/2]$ is the squeezing operator, and $s=|s|e^{i\phi_s}, \phi_s$ are respectively the squeezing parameter and its angle, which depends on the protocol duration $2\tau$.  Although the dynamics under $\hat{H}(t)$ generates squeezing, its amount may in general depend on the initial state. However, for the particular case of thermal states $\hat{\rho}_\beta$ the acquired squeezing does not depend on the initial temperature $\beta^{-1}$. We refer to~\ref{app:dyn} for the details of the derivation. The expression in (\ref{eq:s}) is obtained in the limit $2\omega \tau\rightarrow \infty$; nevertheless, it gives account of the resulting squeezing to a very good approximation for any cycle of duration $2\omega\tau\gtrsim 10$~\cite{Abah:22} and for any initial thermal state. Note also that, as shown in Ref.~\cite{Abah:22}, finite-size systems comprising all-to-all interactions closely follow the expected squeezing given in Eq.~(\ref{eq:s}).

\begin{figure}
 \centering
 \includegraphics[width=0.8\linewidth,angle=-0]{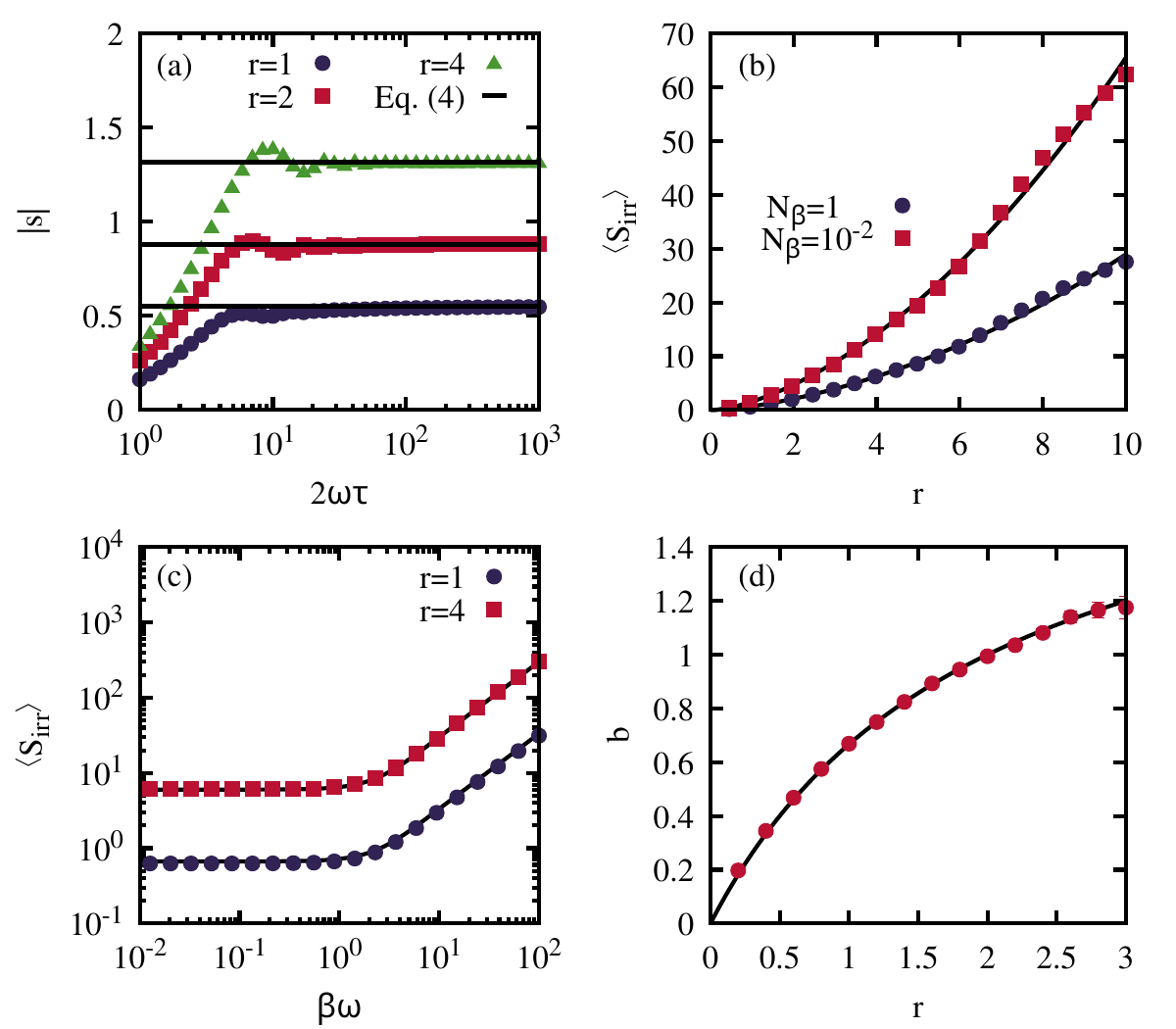}
 \caption{\label{fig1}\small{(a) Squeezing generated upon the completion of a closed cycle with $g_f=g_c$, as a function of the cycle duration $2\omega\tau$. Solid points correspond to the exact dynamics (solved numerically) of the system that is initialized in an initial thermal state $\hat{\rho}_\beta$, and driven by a nonlinear ramp with different values of $r$, namely, with $r=1$, $2$ and $4$. The solid horizontal lines correspond to the theoretical prediction of $|s|$ as given in Eq.~(\ref{eq:s}), to which $|s|$ saturates.  Although the results have been obtained for $N_\beta=1$, we note that the squeezing $|s|$ is independent of the temperature of the initial state (see~\ref{app:dyn} for details), and thus equivalent results can be found for any other $N_\beta$. Panels (b) and (c) show the irreversible entropy $\langle S_{\rm irr}\rangle/\omega$ at the end of the cycle, as a function of the nonlinear exponent $r$ and the initial inverse temperature $\beta$, respectively. Here, points correspond to exact numerical simulations with $\omega\tau=20$ and solid lines to the theoretical expression in Eq.~(\ref{eq:Sirr_r}). Note that the saturation at high temperature is given by the relation $\langle S_{\rm irr}\rangle = 2\omega \cot^{2}[\pi/(2+2z\nu r)]$. (d) Numerically-fitted exponent from $\langle W_{\rm irr}\rangle-\langle W_{\rm irr}(\tau)\rangle\propto \tau^{-b}$ for an initial quantum vacuum state, with $\langle W_{\rm irr}(\tau)\rangle$ obtained from exact numerical simulations using Eq.~(\ref{eq:H0}) for $\omega\tau\in[10^2,10^4]$, and $\langle W_{\rm irr}\rangle$ given by the theoretical expression in Eq.~(\ref{eq:Sirr_r}). The solid line is $b=2z\nu r/(1+z\nu r)$.}}
\end{figure}

As an example, we show in Fig.~(\ref{fig1})(a) the acquired squeezing upon a complete cycle initializing the system in a thermal state at temperature $\beta^{-1}$ with thermal excitations $N_\beta$ as a function of the cycle duration $2\omega \tau$ and for three different $r$ exponents. Although the results shown in Fig.~\ref{fig1}(a) have been obtained for $N_\beta=1$, we stress that the resulting squeezing is independent of $\beta$.  The closed dynamics are dictated by the von Neumann equation, $\partial_t\hat{\rho}(t)=-i[\hat{H}(t),\hat{\rho}(t)]$, and can be easily solved employing Gaussian states as the Hamiltonian in Eq.~(\ref{eq:H0}) is quadratic in the bosonic operators $\hat{a}$ and $\hat{a}^\dagger$~\cite{Ferraro:05} (see~\ref{app:dyn} for details). Our numerical results show that the analytical expression in Eq.~(\ref{eq:s}) corresponds to the squeezing obtained for $2\omega\tau\gtrsim 10$. For rapid cycles, the initial state remains trivially unaltered, so that $\hat{\rho}(2\tau)\approx \hat{\rho}_\beta$ holds for $2\omega\tau\ll 1$. In the following, we will focus on the nontrivial regime where the cycle is performed in a timescale larger than the inverse of the natural frequency of the system $\omega^{-1}$, while we refer to~\ref{app:transient} for a discussion of the dynamics in the transient regime $1\lesssim 2\omega \tau\lesssim 10$. 

The generation of squeezing stems from nonadiabatic excitations promoted due to the QPT at $g_c$, which is more pronounced as the nonlinear exponent $r$ increases because the system expends more time close to the critical point. Indeed, from Eq.~(\ref{eq:s}) it follows that $|s|\approx \log[4z\nu r/\pi]$ for $r\gg 1$, so that the amount of squeezing diverges in the limit $r\rightarrow \infty$. To the contrary,  $|s|\rightarrow 0$ for $r\rightarrow 0$, since in such a limit the control transformation remains constant (i.e., $g(t)=0$) for $t\in[0,2\tau]$ and the system is not brought to the QPT. It is worth noting that, although the generated squeezing is given by Eq.~(\ref{eq:s}) to a very good approximation for slow cycles, there are finite-time corrections that may be significant depending on the nonlinear exponent $r$. We refer to Sec.~\ref{s:Sirr} for a discussion on how these finite-time corrections manifest in the system.

Before moving forward to analyze relevant thermodynamic quantities, such as work and irreversible entropy, we observe that a squeezed thermal state (as for example the final state  $\hat{\rho}(2\tau)$) becomes nonclassical when $N_\beta\leq (e^{2|s|}-1)/2$, since the $P$-diagonal representation of the state is negative or diverging~\cite{Kim:89}. In the driven mean-field critical model, the inequality $N_\beta\leq (e^{2|s|}-1)/2$ translates to a condition between the generated squeezing and the inverse temperature $\beta$ of the initial state, i.e.~$|s|\geq \log[\cot[\beta \omega/2]]/2$. Hence, using Eq.~(\ref{eq:s}), we can determine the threshold nonlinear exponent $r_c$ above which the initial thermal state $\hat{\rho}_\beta$ is squeezed, by applying a cycle that reaches the critical point $g_f=g_c$, and yielding at the same time nonclassical properties: 
\begin{equation}
   r_c = -2+\pi \ {\rm arcsin}^{-1}\left[ \frac{ \sqrt{1+2N_\beta} }{ (1+N_\beta) }\right].
\end{equation}  
As an example, for $N_{\beta}=1$, i.e.~$\beta \omega=\log[2]$, this is achieved using $g(t)$ with $r\geq r_c=1$. As the thermal occupation grows, the cycle to guarantee nonclassicality, in terms of the $P$-diagonal representation of the state, must be realized with a larger nonlinear exponent, which increases the generated squeezing. This is because $r_c \approx \pi\sqrt{N_\beta/2}-2$ for $N_\beta\gg 1$. 
 
\section{Work statistics and Irreversible entropy}\label{s:Sirr}

Let us start with the definitions of internal energy, work, and heat for the quantum process under scrutiny. From the instantaneous internal energy $\langle U(t)\rangle ={\rm Tr}[\hat{\rho}(t)\hat{H}(t)]$, we find $\langle \partial_t U(t)\rangle =\langle \partial_t Q(t)\rangle +\langle \partial_t W(t)\rangle$ where the heat and work rates are defined as $\langle \partial_t Q(t)\rangle = {\rm Tr}[ \partial_t\left( \hat{\rho}(t) \right)\hat{H}(t) ]$ and $\langle \partial_t W(t)\rangle={\rm Tr}[ \hat{\rho}(t)\partial_t ( \hat{H}(t) ) ]$, respectively. Since we are interested in the coherent dynamics entailed by $\partial_t\left( \hat{\rho}(t)\right) = -i[\hat{H}(t),\hat{\rho}(t)]$, it holds that $\langle \partial_t Q\rangle=0$ for any time $t$. This means that all the changes in the internal energy are due to the work done on the system as a consequence of the protocol $g(t)$, cf.~Eq.~(\ref{eq:gt}).

As aforementioned, the initial thermal state acquires squeezing after a cycle is completed. The work done on the system is a stochastic variable. Being the initial state thermal, and thus without quantum coherence respect to the eigenbasis of $\hat{H}(0)$, we can make use of the two-point measurement scheme~\cite{Esposito:09,Campisi:11} to characterize the work probability distribution $P(W)$, which can capture relevant critical features of the model of interest~\cite{Dorner:12, Funo:17, Mzaouali:21, Zawadzki:20, Varizi:20}. Considering the spectral decomposition $\hat{H}(0)=\sum_n E_n \ket{\varphi_n}\!\bra{\varphi_n}$, the distribution $P(W)$ can be expressed as
\begin{equation}\label{eq:PW}
     P(W) = \sum_{n,m} p_n^0 p_{m|n}^{2\tau} \delta[W-(E_m'-E_n)],
\end{equation}
where $p_n^0$ denotes the probability of measuring the state in the eigenstate $\ket{\varphi_n}$ with energy $E_n$ at time $t=0$, and $p_{m|n}^{2\tau}$ is the probability of finding the eigenstate $\ket{\varphi_m'}$ with energy $E_m'$. Here, the eigenstate $\ket{\varphi_m'}$ and energy $E_m'$ relate to the Hamiltonian $\hat{H}(2\tau)=\sum_m E_m'\ket{\varphi_m'}\!\bra{\varphi_m'}$ after completing the protocol conditioned on the initial outcome $E_n$. In our case-study, $\hat{H}(0) = \hat{H}(2\tau) = \omega \, \hat{a}^\dagger \hat{a}$ whose eigenstates are the Fock states $\ket{n}$ such that $\hat{a}^\dagger \hat{a}\ket{n}=n\ket{n}$. More precisely, $p_n^0=\bra{n}\hat{\rho}(0)\ket{n}$ and $p_{m|n}^{2\tau}=\bra{m} \hat{\mathcal{U}}\ket{n}\bra{n} \hat{\mathcal{U}}^\dagger \ket{m}=|\bra{m}\hat{\mathcal{U}}\ket{n}|^2$ where $\hat{\mathcal{U}}=\hat{\mathcal{T}} \exp[-i\int_0^{2\tau} dt \hat{H}(t)]$ denotes the time-evolution propagator.

Let us now analyze the irreversible work. The mean value of the total work is equal to 
\begin{equation}
 \langle W\rangle={\rm Tr}\left[ \hat{\rho}(2\tau) \hat{H}(2\tau) \right] - {\rm Tr}\left[ \hat{\rho}(0)\hat{H}(0)\right], 
\end{equation}
and the irreversible work reads as $\langle W_{\rm irr}\rangle=\langle W\rangle -\Delta F$, where $\Delta F = F(2\tau) - F(0)$ is the difference between final and initial equilibrium free energies with $F(t) = -\beta^{-1}\log{\rm Tr}[e^{-\beta \hat{H}(t)}]$. Since we are studying a closed cycle, this difference vanishes ($\Delta F=0$), meaning that the irreversible work coincides with $\langle W\rangle$: $\langle W_{\rm irr}\rangle = \langle W\rangle$. Since $\hat{\rho}(2\tau)$ is a squeezed thermal state, we are left with a simple expression for the irreversible work, i.e.,
\begin{equation}\label{eq:Wirr}
 \langle W_{\rm irr}\rangle=\langle W\rangle=\omega (2N_\beta+1)\sinh^2\big[ |s| \big],
\end{equation}
where we used $\hat{\rho}(0)=\hat{\rho}_\beta$, $\hat{\rho}(2\tau)=\hat{\mathcal{S}}(s)\hat{\rho}_\beta \hat{\mathcal{S}}^\dagger(s)$ and $N_\beta=(e^{\beta \omega}-1)^{-1}$. As expected, states at larger temperatures produce larger irreversible work since $\langle W_{\rm irr}\rangle$ grows linearly with $N_\beta$. 

Eq.~(\ref{eq:Wirr}) allows us to analyze also the average irreversible entropy $\langle S_{\rm irr}\rangle = \beta \langle W_{\rm irr}\rangle$, which is responsible for a reduction of extractable work~\cite{MohammadyCP2020}. In this regard, it is important to clarify that the thermodynamic cycle that gives rise to the squeezed thermal state $\hat{\rho}(2\tau)$ is a closed driving protocol. Once the unitary cycle is completed at time $t=2\tau$, we assume that the system is coupled to a thermal bath with inverse temperature $\beta$, as in Refs.~\cite{Varizi:20, Santos:19}. Due to the macroscopic nature of the thermal bath, the state of the system tends in a very short time towards the thermal state $\exp(-\beta \hat{H}(2\tau))/{\rm Tr}[\exp(-\beta \hat{H}(2\tau))]$, whereby the resulting entropy corresponds precisely to $\langle S_{\rm irr}\rangle$. Within this perspective, we study $\langle S_{\rm irr}\rangle$ and refer to it as an irreversible entropy even for the unitary evolution examined in our study. Depending on the initial temperature $\beta^{-1}$, we find two limiting regimes for $\langle S_{\rm irr}\rangle$:
\begin{equation}\label{eq:Sirr}
 \langle S_{\rm irr}\rangle =\beta \langle W_{\rm irr}\rangle =
 \Bigg\lbrace
 \begin{array}{cc}
 2\sinh^2\big[ |s| \big]&{\rm for}\;\beta\omega\ll 1,\\ 
 \beta \omega \sinh^2\big[ |s| \big]&{\rm for}\; \beta\omega\gg 1. 
 \end{array}
\end{equation}
In the high-temperature limit $\beta\omega\rightarrow 0$, the average irreversible entropy saturates to the value $\langle S_{\rm irr}\rangle =2\sinh^2[|s|]$, while the irreversible work diverges. Conversely, in the zero-temperature limit $\beta\omega\rightarrow \infty$, the irreversible work is $\langle W_{\rm irr}\rangle =\omega \sinh^2[|s|]$ and corresponds to the energy of excitations produced during the finite-time cycle, with a diverging $\langle S_{\rm irr}\rangle$. 

We recall that, in the physical context analyzed in this paper, the average irreversible entropy can be expressed also as~\cite{Deffner:10} $\langle S_{\rm irr}\rangle=S(\hat{\rho}(2\tau)||\hat{\rho}_{\rm eq})$ where $S(x||y)={\rm Tr}[x\log x-x\log y]$ denotes the quantum relative entropy, and $\hat{\rho}_{\rm eq}$ is the equilibrium state at the end of the protocol at temperature $\beta^{-1}$. In our case, $\hat{\rho}_{\rm eq}$ is the initial thermal state $\hat{\rho}_\beta$, and $\Delta F = 0$. We will come back to this in Sec.~\ref{s:coh} to assess the role of quantum coherence, originated by the thermodynamic cycle, in the irreversible entropy.

So far, we have expressed the irreversible work and entropy in terms of the acquired squeezing amplitude $|s|$. Now let us analyze the case when the cycle does not reach the critical point, i.e.~$g_f<g_c$. In this scenario, the adiabatic condition applies for $\omega\tau\gg 1$ since the energy gap is non-zero during the total evolution. Hence, for sufficiently slow cycles the initial state is recovered, that is, $|s|\rightarrow 0$ for $2\omega\tau\rightarrow \infty$ and similarly $\langle W_{\rm irr}\rangle\rightarrow 0$. To the contrary, when the cycle reaches the quantum critical point for $g_f=g_c=1$ in Eq.~(\ref{eq:gt}), the squeezing amplitude $|s|$ is solely determined by both the critical exponents and $r$ to a very good approximation for any protocol duration provided $2\omega\tau\gtrsim 10$. In such a case, irreversible work and entropy inherit the dependence on the critical exponent and are determined by a universal function, dependent only on $\beta$ and $z\nu r$. In particular, we find that
 \begin{equation}\label{eq:Sirr_r}
     \langle S_{\rm irr}\rangle=\beta \langle W_{\rm irr}\rangle=\beta\omega \coth\left[\frac{\beta\omega}{2}\right]\cot^{2}\left[\frac{\pi}{2(1+z\nu r)}\right].
 \end{equation}
 From Eq.~(\ref{eq:Sirr_r}), it follows that the average irreversible entropy saturates to $\langle S_{\rm irr}\rangle = 2\cot^{2}[\pi/(2(1+z\nu r))]$ in the high-temperature limit $\beta\omega \rightarrow 0$, and increases as $\langle S_{\rm irr}\rangle = \beta\omega \cot^2[\pi/(2(1+z\nu r))]$ as the initial temperature goes to zero ($\beta\omega\rightarrow \infty$). In Fig.~(\ref{fig1}), panels (b) and (c), we show the average irreversible entropy as a function of the initial inverse temperature $\beta$ and the exponent $r$ respectively, and we compare the analytical expression (\ref{eq:Sirr_r}) with the exact numerical results obtained for $2\omega\tau=20$. It can be observed that the larger the value of $r$, which controls how the system approaches the critical point, the more irreversible is the cycle. This trend gets more pronounced for growing $\beta\omega$, also because the irreversible work starts losing the dependence on $\beta$ when the temperature of the initial thermal state tends to zero. Note that for a fixed $r$, the behavior of $\langle W_{\rm irr}\rangle$ and $\langle S_{\rm irr}\rangle$ as a function of $\beta\omega$ only depends on the initial thermal occupation, which enters as a prefactor (cf. Eqs.~(\ref{eq:Wirr}) and~(\ref{eq:Sirr_r})) given that the generated squeezing is agnostic to the initial temperature of the state. These results are a consequence of the breakdown of the adiabatic condition in a zero-dimensional model~\cite{Polkovnikov:05, Polkovnikov:08}, cf.~Eq.~(\ref{eq:H0}). Even for a cycle performed infinitely slow ($2\omega\tau\rightarrow \infty$), the irreversible work $\langle W_{\rm irr}\rangle$ remains non-zero and constant in agreement with the results reported in Ref.~\cite{Gherardini:24}. In contrast, for non-zero dimensional systems  $\langle W_{\rm irr}\rangle \propto \tau^{-d\nu r/(1+z\nu r)}$ as successfully described by the Kibble-Zurek mechanism~\cite{delCampo:14, Zurek:05, Barankov:08, Sen:08}. We refer to~\ref{app:Ising} for a comparison with the paradigmatic transverse-field Ising model in one dimension, where $\langle W_{\rm irr}\rangle$ decreases as $2\omega\tau$ increases as dictated by the Kibble-Zurek scaling. The absence of a saturation in the irreversible work for low-dimensional systems, such as in the transverse Ising model, highlights the different behavior with respect to mean-field critical systems that follow the trend given by Eq.~(\ref{eq:Sirr_r}).

 As commented previously and shown in Fig.~(\ref{fig1})(a), finite-time corrections to the expressions derived for an infinitely slow cycle are expected. Although these corrections are small for $2\omega\tau\gtrsim 10$, they may still be significant depending on the parameters considered. As an example, we take into account the average irreversible work. In particular, we find that $\langle W_{\rm irr}\rangle-\langle W_{\rm irr}(\tau)\rangle\propto \tau^{-b}$, where $b$ depends on the nonlinear exponent $r$, and $\langle W_{\rm irr}\rangle$ and $\langle W_{\rm irr}(\tau)\rangle$ are respectively the theoretical values of the average irreversible work given by Eq.~(\ref{eq:Sirr_r}) and the result of the numerical simulations for a finite driving time $\tau$. For a driving function stopping at the critical point, it is known that the excess energy, corresponding to the average work done on the system, scales following the Kibble-Zurek mechanism, i.e.~$\langle W\rangle\sim \tau^{-z\nu r/(1+z\nu r)}$~\cite{Hwang:15}. However, for a cycle $\langle W\rangle$ tends to be constant for large $\tau$, but its finite-time correction for infinitely slow cycles still scales with $\tau$ in a universal manner as $\tau^{-b}$ with $b=2z\nu r/(1+z\nu r)$. Interestingly, this corresponds to twice the Kibble-Zurek scaling exponent~\cite{deGrandi:10c, deGrandi:10, deGrandi:10b, Hwang:15}. In Fig.~(\ref{fig1})(d) we show the numerically-fitted exponent $b$ for an initial quantum vacuum state as a function of $r$. A similar scaling for the fitted exponent $b$ holds for any $N_\beta>0$. The fits have been performed in the range $\omega\tau\in [10^2,10^4]$, while the solid line corresponds to  $b=2z\nu r/(1+z\nu r)$.

\begin{figure}[t!]
 \centering
 \includegraphics[width=0.7\linewidth,angle=-0]{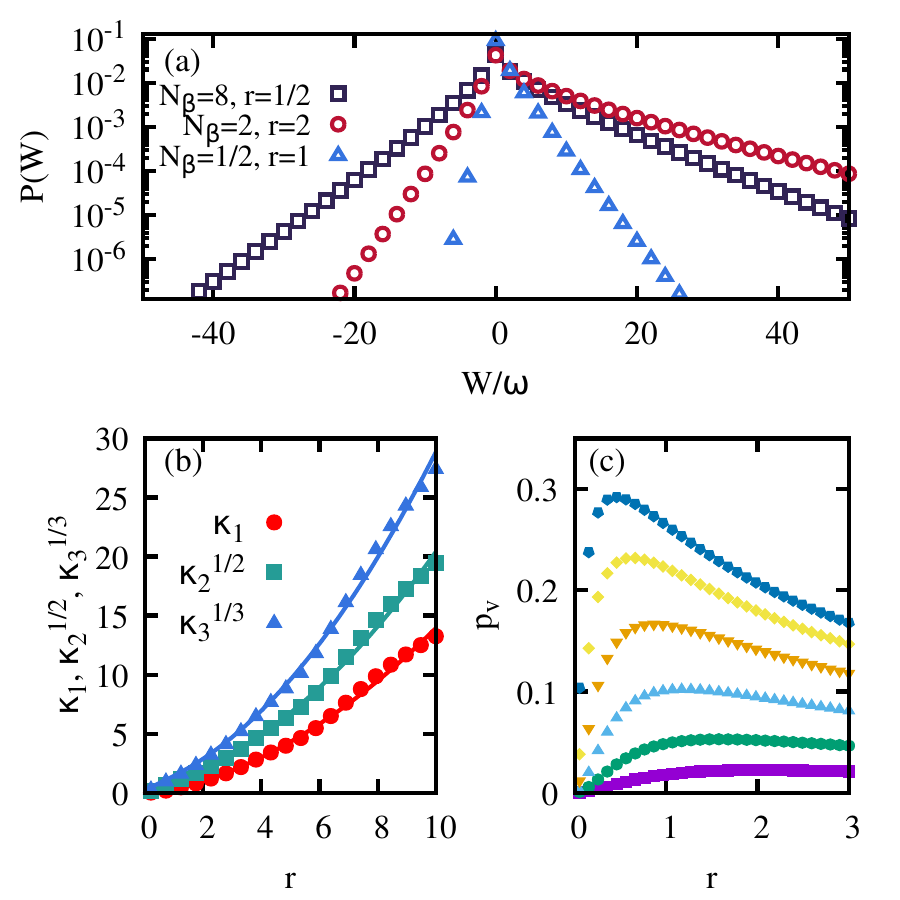}
 \caption{\label{fig2}\small{(a) Work probability distributions $P(W)$ upon a cycle with $2\omega\tau=20$ and different initial states and nonlinear exponents as stated in the legend. (b) First three cumulants of $P(W)$ for $\beta\omega\rightarrow\infty$. The points correspond to the results of exact numerical simulations with $\omega\tau=20$, while the solid lines refer to the universal expressions given in the main text, solely determined by the critical exponents $z\nu$ and $r$. (c) Probability $p_v = \int_{W<0}dW P(W)$ of observing events where a negative work is done after the cycle, plotted as a function of $r$ for increasing amounts of thermal excitations from bottom to top points, $N_\beta=0.5$ (magenta), $1$, $2$, $4$, $8$ and $16$ (blue)}.}
 \end{figure}

 Let us now turn our attention to the statistics of the work distribution $P(W)$. Thanks to the simplicity of the initial and final states, we can exactly compute $P(W)$ and its cumulants (see also Ref.~\cite{Deffner:08} for a related study). Since the time-evolution produces squeezing,  $\hat{\mathcal{U}}=\hat{\mathcal{S}}(s)$,  it follows that $p_{m|n}^{2\tau} = \left| \bra{m}\hat{\mathcal{S}}(s)\ket{n} \right|^2$, while $p_n^0$ is the population of the $n$-th Fock state in the thermal state: $p_n^0=N_\beta^n(1+N_\beta)^{-(n+1)}$. The quantity $p_{m|n}^{2\tau}$ is the population associated to a number state $\ket{n}$ squeezed on $\ket{m}$, whose expression can be analytically computed~\cite{Kim:89}. To make the connection with $\hat{\mathcal{S}}(s)$ more explicit, we denote $p_{m|n}^{2\tau}$ as $S_{n,m}(|s|)$; see~\ref{app:ST} for details. It is worth noting that the population only depends on the modulus of the squeezing parameter. Moreover, due to the structure of the Hamiltonian $\hat{H}(0) = \hat{H}(2\tau)=\omega \, \hat{a}^\dagger \hat{a}$, the work probability distribution $P(k\omega)=0$ for $k$ odd; this can also be deduced from the fact that $p_{m|n}^{2\tau}=0$ if $m-n$ is odd. Therefore, we can rewrite the work probability distribution as
 \begin{equation}
  P(2k\omega)=\sum_{n\geq 0}\sum_{m=2k+n,\,m\geq 0} \frac{N_\beta^n}{(1+N_\beta)^{n+1}}S_{n,m}(|s|)
 \end{equation}
 with $k\in \mathbb{Z}$. If $g_f = g_c$, the probability distribution of work is independent of $\tau$ to a very good approximation provided $2\omega\tau\gtrsim 10$, so that $|s|$ is given by Eq.~(\ref{eq:s}). Hence, $P(W)$ solely depends on $r$ and the critical exponents $z\nu=1/2$. Although one can compute exactly the previous distribution for any initial thermal state and nonlinear exponent $r$, the resulting expressions are intricate. Thus, we illustrate $P(W)$ in Fig.~(\ref{fig2})(a) for different initial thermal states and nonlinear exponents.

 Analytical expressions for the cumulants of the work probability distribution can be provided when the initial state is the vacuum, corresponding to the limit $\beta\omega\rightarrow \infty$. For $\hat{\rho}(0)=\ket{0}\!\bra{0}$ as in Ref.~\cite{Gherardini:24}, the first and second cumulants are $\kappa_1=\langle W\rangle = \omega \sinh^2[|s|]$ and $\kappa_2=(\Delta W)^2=\langle W^2\rangle -\langle W\rangle^2=2\omega^2\cosh^2[|s|]\sinh^2[|s|]$, respectively, which describe the mean value and variance of the distribution. The third cumulant, $\kappa_3=\int dW P(W)(W-\langle W\rangle)^3=\omega^3\cosh[2|s|]\sinh^2 [2|s|]$, is related to the skewness of $P(W)$, which can be quantified utilizing the third standardized moment $\tilde{\mu}_3=\kappa_3/\kappa_2^{3/2}=2\sqrt{2}\cot[2|s|]$. Again when $g_f=g_c$, the cumulants of the work probability distribution acquire a universal form depending only on the critical properties of the model (cf.~Fig.~\ref{fig2}(b) for a representation of $\kappa_{1,2,3}$ as a function of $r$, while we refer to~\ref{app:transient} for the behavior of $\kappa_3$ for $N_\beta> 0$ and different $r$). Moreover, $\tilde{\mu}_3\approx 2\sqrt{2}$ for $r\gg 1$, while it diverges as $\tilde{\mu}_3\propto 1/(z\nu r)$ for $r\ll 1$. As expected, the distribution $P(W)$ features a positive skewness $\tilde{\mu}_3$ since the negative values are exponentially suppressed by a factor $e^{-\beta W}$, cf.~Fig.~(\ref{fig2})(a), which follows from the Tasaki-Crooks theorem $P(-W)=P(W)\exp[-\beta W]$~\cite{Crooks:99}.

 Due to quantum and thermal fluctuations, the second law is only satisfied on average, $\langle S_{\rm irr}\rangle\geq 0$ [cf.~Eq.~(\ref{eq:Sirr_r})]. At the stochastic level, since $S_{\rm irr}=\beta W_{\rm irr}$, an event violating the second law corresponds to $W_{\rm irr}<0$. Thus, the probability $p_v=\int_{W<0} dW P(W)$ quantifies how likely is to observe $S_{\rm irr}<0$. The probability $p_v$ exhibits a maximum for a particular nonlinear exponent $r$, which depends on the initial thermal occupation number $N_\beta$. As commented before, the irreversibility is enhanced for large values of $r$ and $\beta\omega$, which thus means a decreasing value of $p_v$. This is illustrated in Fig.~(\ref{fig2})(c). However, for larger temperatures $\beta\omega\rightarrow 0$, the probability $p_v$ increases given that $P(W)$ and $P(-W)$ become more comparable among them such that $P(-W)/P(W) = e^{-\beta W} \approx 1$. Yet, as the temperature grows, a small amount of squeezing generated during the cycle (corresponding to a small $r$) is enough to make events leading to $S_{\rm irr}<0$ more probable. In particular, $p_v$ is maximum at $r\approx 1/2$ for $N_\beta=16$, and at $r\approx 1.2$ for $N_\beta=2$, while for an initial vacuum state ($\beta\omega\rightarrow \infty$), $p_v=0$ regardless $r$ since $P(k\omega)=0 \ \forall k<0$~\cite{Gherardini:24}. Therefore, the probability to observe events where the second law seems to be violated, plotted as a function of $r$, follows an opposite trend when compared with the behavior of the final state that is more nonclassical as the value of $r$ increases. Indeed, a squeezed thermal state becomes nonclassical for a sufficiently large squeezing parameter, and in our case, this translates to having $r\geq r_c$, with $r_c$ increasing at higher temperatures (cf.~Sec.~\ref{s:model}).
  
 \begin{figure}[t!]
 \centering
 \includegraphics[width=0.85\linewidth,angle=-0]{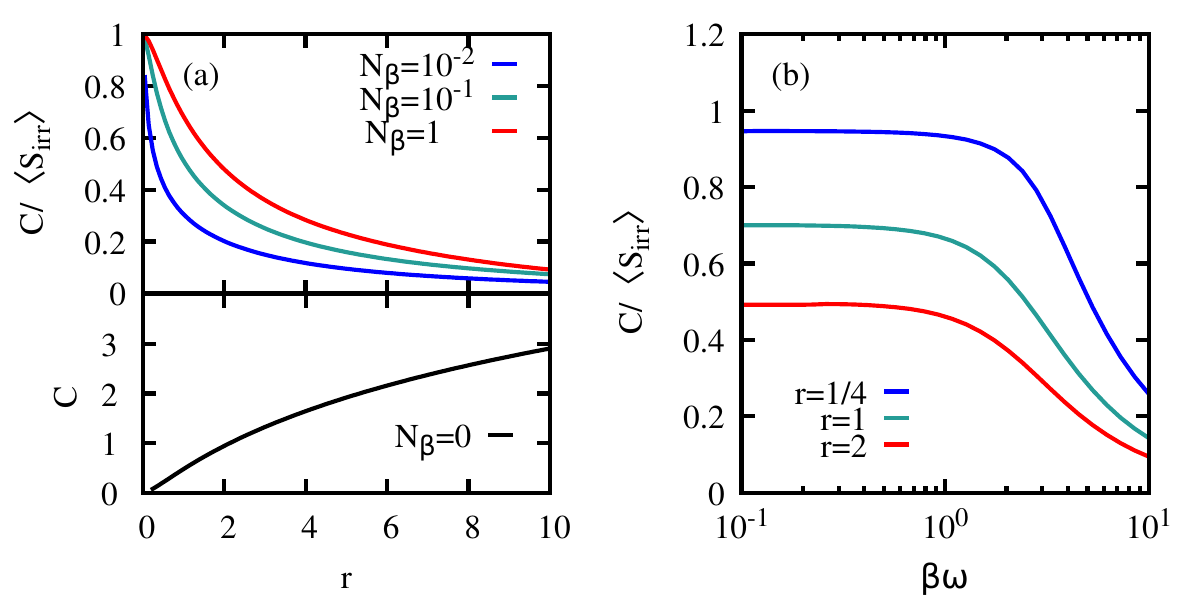}
 \caption{\label{fig3}\small{Contribution from the coherence to the average irreversible entropy quantified through the ratio $C/\langle S_{\rm irr}\rangle$, shown as a function of the nonlinear exponent $r$ for different initial occupations $N_\beta$ [top plot in panel (a)], and $C$ for the zero-temperature limit [bottom plot in panel (a)]. Panel (b) shows the ratio $C/\langle S_{\rm irr}\rangle$ as a function of the initial inverse temperature $\beta$ for distinct $r$ values.}} \end{figure}

\section{Coherence contribution to the irreversible entropy}\label{s:coh}

The average irreversible entropy can be divided into two contributions~\cite{Santos:19, Varizi:20}, that is, 
 \begin{equation}\label{eq:splitting}
 \langle S_{\rm irr}\rangle = D + C,
 \end{equation}
 where $D = S( \Delta[\hat{\rho}(2\tau)] \, || \, \hat{\rho}_\beta )$ is the contribution due to changes in populations, and $C$ is due to the quantum coherences created by the nonequilibrium process. The latter is quantified in terms of the relative entropy of coherence, and reads as~\cite{Santos:19, Varizi:20}
 \begin{equation}
 C=S_v\left( \Delta[\hat{\rho}(2\tau)] \right) -S_v\left( \hat{\rho}(2\tau) \right),
 \end{equation}
 where we have introduced the von Neumann entropy $S_v(\hat{\rho})=-{\rm Tr}[\hat{\rho}\log[ \hat{\rho}]]$ and $\Delta[\hat{\rho}]$ that denotes the completely dephased state (i.e., with no coherence) in the eigenbasis of the final Hamiltonian.  As discussed in Ref.~\cite{Varizi:21}, it is worth noting that the splitting in Eq.~(\ref{eq:splitting}) may not properly distinguish coherent from non-coherent driving protocols, but describes how populations and off-diagonal terms contribute to entropy production on average. This is also due to a divergent trend of $\langle S_{\rm irr}\rangle$ for $\beta \rightarrow \infty$. In the following, we will address this issue by focusing our attention on the relative entropy of coherence alone, which is not a divergent quantity and depends proportionally on the squeezing generation.

 We are interested in determining the role that quantum coherences originating from the dynamics play in the average irreversible entropy. Since the evolution is unitary, $\hat{\rho}(2\tau)=\hat{\mathcal{U}}\hat{\rho}(0)\hat{\mathcal{U}}^\dagger$, the von Neumann entropy of the final state is identical to that of the initial thermal state: $S_v(\hat{\rho}_\beta)=\beta \exp[\beta] N_\beta+\log [N_\beta]$. On the other hand, the state $\hat{\rho}(2\tau)$ is a squeezed thermal state; thus, the completely dephased state in the Fock basis is
  \begin{equation}
    \Delta\left[ \hat{\rho}(2\tau) \right] = \Delta\left[ \hat{\mathcal{S}}(s)\hat{\rho}_\beta \hat{\mathcal{S}}^\dagger(s) \right] = \sum_{n=0}^\infty p_n(\beta,s)\ket{n}\!\bra{n},
 \end{equation}
  where $\sum_n p_n(\beta,s)=1$. The probabilities $p_n(\beta,s)$ can be computed exactly [cf.~\ref{app:ST}]. Accordingly, the relative entropy of coherence is
  \begin{equation}
  C=-\beta \exp[\beta] N_\beta-\log[N_\beta]-\sum_{n=0}^\infty p_n(\beta,s)\log[p_n(\beta,s)],
  \end{equation}
  which quantifies the contribution to $\langle S_{\rm irr}\rangle$ due to the quantum coherence generated during the cycle. 

 Employing the squeezing $|s|$ given in Eq.~(\ref{eq:s}), we can compute the ratio $C/\langle S_{\rm irr}\rangle$, belonging to the interval $[0,1]$, for any nonlinear exponent $r$ and inverse temperature $\beta$. The results are plotted in Fig.~(\ref{fig3}). Let us discuss the limit cases. 

 The ratio $C/\langle S_{\rm irr}\rangle$ tends to the maximum value $1$ for cycles performed with decreasing nonlinear exponents $r$ ($r\lesssim 1$), considering an initial thermal state at high temperature. In this case, most of the irreversible entropy stems from the coherence contribution, which is consistent with previous works in non-zero dimensional systems, such as spin chains~\cite{Varizi:20}. This is illustrated in Fig.~\ref{fig3}(b). Indeed, in the limits $r,\beta\omega\ll 1$, we find that $C\approx \langle S_{\rm irr}\rangle\approx \pi^2 (z\nu r)^2/2$ that however takes quite small values. In the opposite limit, when  $\beta\omega\rightarrow \infty$ (zero temperature limit), it follows that $S_v(\hat{\rho}(2\tau))=0$ since the final state is pure, $\hat{\rho}(2\tau)=\hat{\mathcal{S}}(s)\ket{0}\!\bra{0}\hat{\mathcal{S}}^\dagger(s)$. Therefore, the relative entropy of coherence $C$ reduces to a Shannon entropy:
 \begin{equation}
  C=-\sum_{n=0}^{\infty}p_n^{\infty}(s)\log [p_n^{\infty}(s)],
 \end{equation}
  where $p_n^{\infty}(s)=\lim_{\beta\rightarrow\infty}p_n(\beta,s)$ is the standard photon distribution of a squeezed vacuum state,
  \begin{equation}\label{eq:photon_distr_vacuum_state}
  p_{2n}^{\infty}(s)=\frac{(2n)!}{4^n(n!)^2}{\rm sech}[|s|] \tanh^{2n}[|s|],
 \end{equation}
  with $p_{2n+1}^{\infty}(s)=0, \ \forall n\geq 0$. Substituting Eq.~(\ref{eq:s}) in (\ref{eq:photon_distr_vacuum_state}), the populations of even Fock states become 
  \begin{equation}
  p_{2n}^\infty(r)=\frac{(2n)!}{4^n(n!)^2}  \sin\left[\frac{\pi}{2+2z\nu r} \right]\cos^{2n}\left[\frac{\pi}{2+2z\nu r} \right]
  \end{equation}
  that follows a negative binomial distribution with a fractional number of successes~\cite{Gherardini:24}, given that $\frac{(2n)!}{4^n(n!)^2} = \frac{\Gamma(n+\frac{1}{2})}{n!\,\Gamma(\frac{1}{2})}$. This allows us to express the relative entropy of coherence $C$ in its universal form. In this regard, it is worth noting that in the zero-temperature limit, $\langle S_{\rm irr}\rangle$ diverges given that $\langle S_{\rm irr}\rangle = \beta \langle W_{\rm irr}\rangle$ with $\langle W_{\rm irr}\rangle$ a constant quantity for $\beta\omega\rightarrow \infty$. For the average irreversible entropy $\langle S_{\rm irr}\rangle$, the relative entropy of coherence $C$ plays a negligible role given that $\lim_{\beta\rightarrow\infty}C/\langle S_{\rm irr}\rangle=0$ [see the top plot of Fig.~(\ref{fig3})(a)], so that $\langle S_{\rm irr}\rangle\approx D$ in such a limit. However, $C$ is independent of $\beta$ and has a finite, monotonic trend with $r$. This trend of $C$ is directly proportional to the generation of squeezing by the dynamics, whose magnitude indeed grows with $r$. We plot $C$ as a function of $r$ in the bottom plot of Fig.~(\ref{fig3})(a) that also indicates $C\rightarrow \infty$ for $r\rightarrow \infty$. We can thus conclude the following: {\rm i)} the relative entropy of coherence $C$ is the non-diverging part of the average irreversible entropy in the limit of zero temperature. {\rm ii)} The generation of squeezing, leading to the formation of excitations that may break the adiabatic theorem, is accompanied by a signature of irreversibility that is well-captured by the relative entropy of coherence $C$.

\section{Conclusions}\label{s:conc}

 In this article, we investigate the nonequilibrium probability distribution of work and irreversible entropy, resulting from a time-dependent driving of a critical system that undergoes a mean-field quantum phase transition. In particular, our analysis focuses on cyclical protocols that drive the system, initialized in a thermal state, from an initial control parameter to its critical value and back to the initial value in a finite time. The driving function is expressed in terms of the nonlinear exponent $r$ that modifies the functional form with which the system approaches the critical point. Provided the cycle is performed slow enough, namely in a time larger than the inverse of its natural frequency, the resulting dynamics generates squeezing, whose amount is solely determined by the critical exponents of the mean-field quantum phase transition, namely $z\nu=1/2$, and the nonlinear exponent $r$.

 Interestingly, the average irreversible work and entropy do not depend on the driving time, in sharp contrast to non-zero dimensional many-body systems. The average irreversible work saturates to a non-zero value even in the limit of infinitely slow cycles. This allows us to obtain analytical expressions in terms of both the temperature of the initial thermal state and the nonlinear exponent $r$ of the driving function. Moreover, the work probability distribution can be computed exactly, allowing for the determination of its cumulants. We show that tuning the nonlinear exponent $r$ can increase the probability of observing events with a negative work done after the cycle, which leads to $S_{\rm irr}<0$. 

 Finally, we investigate the quantum coherence created during the non-equilibrium dynamics of the system and determine its role in the irreversible entropy production. For high-temperature states, the average irreversible entropy mostly stems from the quantum coherence generated by cycles where the control is varied with a nonlinear exponent $r$ small in magnitude; since it scales as $r^2$, $\langle S_{\rm irr}\rangle$ can be considered negligible. On the contrary, for low-temperature states ($\beta\omega\rightarrow \infty$), the irreversible work no longer depends on the temperature, and the average irreversible entropy $\langle S_{\rm irr}\rangle$ is thus a divergent quantity. However, we determine that the generation of squeezing entailed by crossing a mean-field critical point has traits of irreversibility that are quantified by the relative entropy of coherence $C$. The latter admits a universal expression that monotonically increases as a function of $r$.

Our results aim to mark an advancement in understanding the complex interplay between critical behaviors and quantum thermodynamics in the broad range of nonequilibrium quantum systems to which our model applies.  
 
 \section*{Acknowledgments}
 F.J.G-R.~gratefully acknowledges financial support from the Spanish MCIN, with funding from the European Union Next Generation EU (PRTRC17.I1), as well as from the Consejer\'ia de Educaci\'on, Junta de Castilla y Le\'on, through the QCAYLE project. Additional support from the Department of Education, Junta de Castilla y Leon, and FEDER funds (CLU-2023-1-05) is also acknowledged.
 S.G.~acknowledges financial support from the PRIN project 2022FEXLYB Quantum Reservoir Computing (QuReCo), and the PNRR MUR project PE0000023-NQSTI funded by the European Union---Next Generation EU.
 R.P.~acknowledges the Ram{\'o}n y Cajal (RYC2023-044095-I) research fellowship  and project TSI-069100-2023-8 (Perte Chip-NextGenerationEU). 

 \appendix

\section{Dynamical equation and squeezing of thermal states}\label{app:dyn}
 The dynamics under the driven quantum harmonic oscillator $\hat{H}(t)$ (cf. Eq.~(\ref{eq:H0})) can be solved exactly for Gaussian states, which are those with a Gaussian Wigner function in the ${\bf X}=(\hat{x},\hat{p})^T$ phase space, with  $\hat{x}=\hat{a}+\hat{a}^\dagger$ and $\hat{p}=i(\hat{a}^\dagger-\hat{a})$~\cite{Ferraro:05}. Here we are interested in the dynamics of a thermal state $\hat{\rho}_\beta$, which is a Gaussian state with zero first moments, $\langle {\bf X}\rangle=(\langle \hat{x}\rangle,\langle \hat{p}\rangle)^T=(0,0)^T$.  Since the Hamiltonian contains only quadratic operators in terms of $\hat{a}$ and $\hat{a}^\dagger$,  the first moments remain zero at all times.  Therefore, the evolved thermal state is uniquely determined by the covariance matrix ${\bf R}$, whose entries are $R_{i,j}=\frac{1}{2}\langle X_i X_j+X_j X_i \rangle-\langle X_i\rangle\langle X_j\rangle$. Note that the number of bosonic excitations is given by $\langle \hat{a}^\dagger\hat{a}\rangle=({\rm Tr}[{\bf R}]-2)/4$.

The von Neumann equation $\partial_t \hat{\rho}(t)=-i[\hat{H}(t),\hat{\rho}(t)]$ translates into a dynamical equation for the relevant covariance matrix elements, $R_{1,1}$, $R_{1,2}$ and $R_{2,2}$, which can be expressed in a compact form as $\vec{R}(t)=(R_{1,1}(t),R_{1,2}(t),R_{2,2}(t))^T$ and obey the ordinary different equation 
 \begin{equation}
     \frac{d}{dt}\vec{R}(t)=M(t) \vec{R}(0), 
 \end{equation}
with
\begin{equation}
M(t)=\left(\begin{array}{ccc}
     0 & -2\omega (g^2(t)-1) & 0\\ 
     -\omega & 0 & -\omega(g^2(t)-1) \\ 
     0 & -2\omega & 0 
     \end{array}
     \right),
\end{equation}
 being $\vec{R}(0)$  the initial condition. Upon integration for the chosen protocol $g(t)$ one finds the final values as $\vec{R}(2\tau)=L(2\tau)\vec{R}(0)$ with $L(2\tau)=\mathcal{T}e^{\int_0^{2\tau}dt'M(t')}$ and $\mathcal{T}$ denoting the time-ordering operator. The elements $\vec{R}(2\tau)$ determine the final covariance matrix ${\bf R}(2\tau)$. As expected,  ${\rm det}[{\bf R}]$ remains constant during this coherent evolution, i.e. $\frac{d}{dt}{\rm det}[{\bf R}]=0$,  given that ${\rm det}[{\bf R}]^{-1/2}$ quantifies the purity of the state~\cite{Ferraro:05}.
 
 As mentioned in the main text, the squeezing generated by the evolution is independent on the temperature of the initial state. This can be shown as follows: For a thermal state $\hat{\rho}_\beta$, we have the initial condition $\vec{R}_\beta(0)=(1+2N_\beta,0,1+2N_\beta)^T$. Consider first a thermal state with zero excitations, $N_\beta=0$ or similarly $\beta\omega\rightarrow\infty$, i.e. an initial vacuum state, so that $\vec{R}_\infty(0)=(1,0,1)^T$. This leads to $\vec{R}_\infty(2\tau)=L(2\tau)\vec{R}_\infty(0)$. By diagonalizing the  covariance matrix one finds ${\rm eig}[{\bf R}_{\infty}]=(e^{2|s|},e^{-2|s|})^T$, which determines   the resulting squeezing amplitude $|s|$ for an initial vacuum state. Now, for $N_\beta>0$, the initial condition can be expressed as $\vec{R}_\beta(0)=(1+2N_\beta)\vec{R}_\infty(0)$. Since the dynamical map $L(2\tau)$ is independent of $\beta$, we find 
 \begin{eqnarray}
 \vec{R}_\beta(2\tau)&=L(2\tau)\vec{R}_\beta(0)=(1+2N_\beta)L(2\tau)\vec{R}_\infty(0),\nonumber\\
 &=(1+2N_\beta)\vec{R}_\infty(2\tau).
 \end{eqnarray}
 That is, each of the entries of the covariance matrix ${\bf R}_\beta(2\tau)$ is multiplied by a factor $(1+2N_\beta)$ with respect to the vacuum state, ${\bf R}_\beta(2\tau)=(1+2N_\beta){\bf R}_\infty(2\tau)$. As a result,  ${\rm eig}[{\bf R}_\beta(2\tau)]=(1+2N_\beta){\rm eig}[{\bf R}_\infty(2\tau)]=(1+2N_\beta)(e^{2|s|},e^{-2|s|})^T$, that is the squeezing for any $N_\beta>0$ matches the one of the vacuum state. The squeezing generated during a cycle does not depend on the initial temperature of the state.  We stress, however, that two distinct non-thermal initial states will, in general, acquire different squeezing for the same evolution under $\hat{H}(t)$.

  \section{Additional numerical results}\label{app:transient}
\begin{figure}
     \centering
     \includegraphics[width=0.9\linewidth]{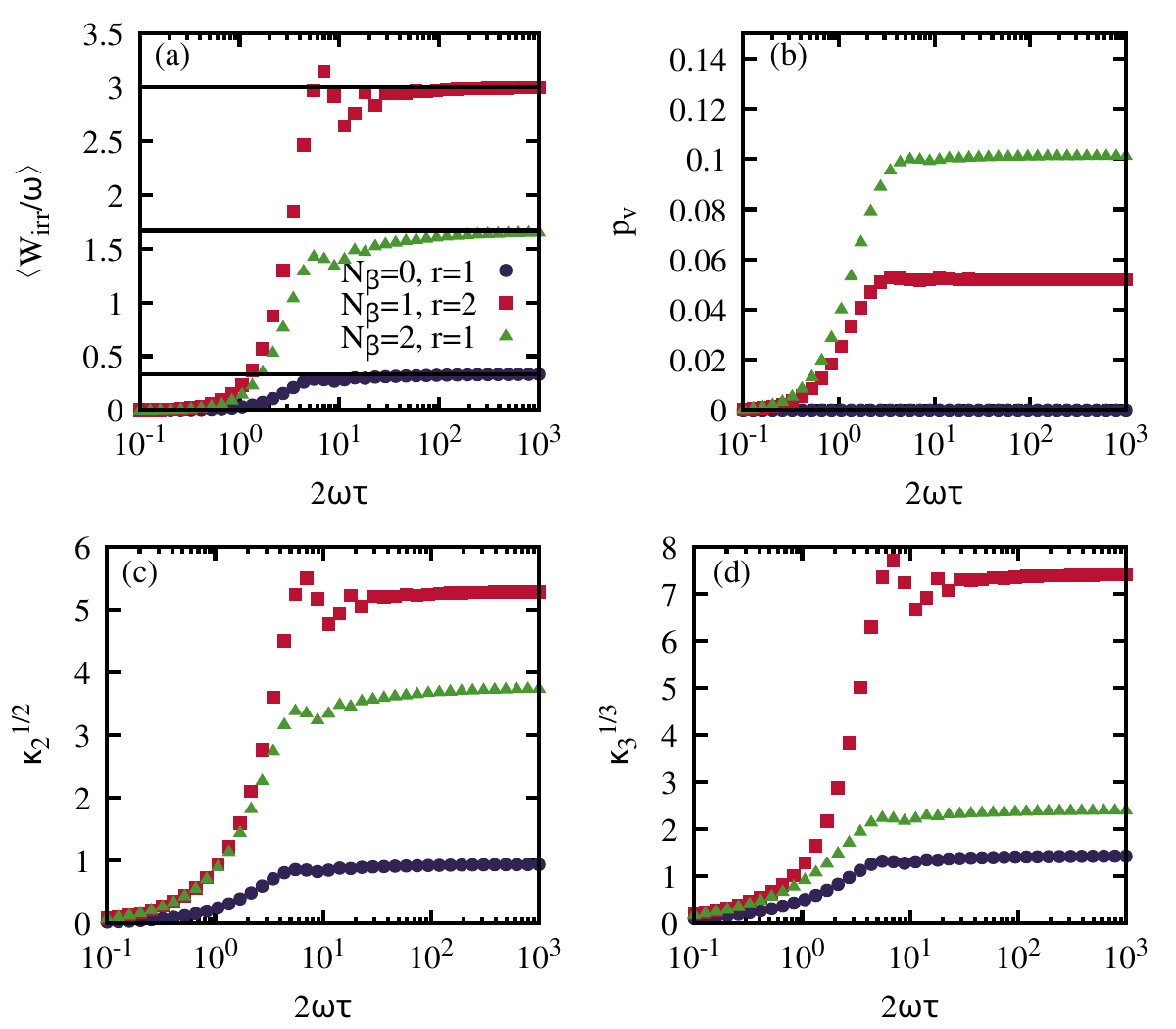}
     \caption{Behavior of the irreversible work (a), probability of negative work done $p_v$ (b), and cumulants $\kappa_2$ (c), $\kappa_3$ (d) of the work distribution $P(W)$ as a function of the cycle duration $2\omega\tau$ for different $r$ exponents and thermal occupations $N_\beta$. Note that the first cumulant $\kappa_1=\langle W_{\rm irr}\rangle$. The solid horizontal lines in panel (a) correspond to the theoretical expression given in Eq.~(\ref{eq:Sirr_r}) for $\langle W_{\rm irr}\rangle$ for the considered $r$ and $N_\beta$ values.}
     \label{fig:transient}
 \end{figure}

 \begin{figure}
     \centering
     \includegraphics[width=0.7\linewidth]{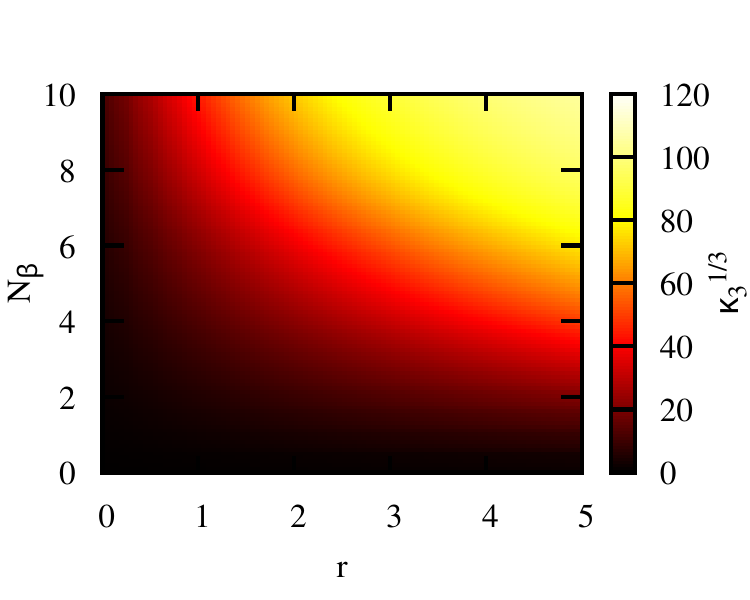}
     \caption{Color map plot of the third cumulant $\kappa_3^{1/3}$ as a function of the exponent $r$ and the number of thermal excitations, $N_\beta=(e^{\beta\omega}-1)^{-1}$. Note that the curve for $N_\beta=0$ is shown in Fig.~\ref{fig2}(b), which is small compared to the growing value of $\kappa_3$ for increasing $N_\beta$ and $r$. The other cumulants, $\kappa_1$ and $\kappa_2^{1/2}$, follow a very similar behavior, being their maximum values in the range shown in this figure approximately $70$, and $90$ in units of $\omega$.} 
     \label{fig:kappa3}
 \end{figure}
 
Here we provide further results regarding the transient regime during which the system acquires a squeezing dependent on the protocol duration $2\omega\tau$, as well as the behavior of the third cumulant $\kappa_3$ for initial states with non-zero temperature.
 
As mentioned in the main text, and illustrated in Fig.~\ref{fig1}(a), the generated squeezing $|s|$ saturates to the theoretical expression given in Eq.~(\ref{eq:s}) for $2\omega\tau\gtrsim 10$, while for very rapid cycles, $2\omega\tau\lesssim 1$, the initial state remains trivially unaltered. For $1\lesssim 2\omega\tau\lesssim 10$, the dynamics find themselves in a transient regime, which has an impact on relevant thermodynamic quantities, such as the work probability distribution $P(W)$, and so their cumulants and the irreversible entropy   $\langle S_{\rm irr}\rangle$. The behavior of these quantities as a function of $2\omega\tau$ is shown in Fig.~\ref{fig:transient}. In particular, we show the first three cumulants of $P(W)$, $\kappa_{1}=\langle W_{\rm irr}\rangle$, $\kappa_2$ and $\kappa_3$, as well as the probability of performing negative work $p_v$, for three different combinations of $r$ and initial thermal occupation $N_\beta$, namely, $N_\beta=0$ and $r=1$, $N_\beta=1$ and $r=2$ and $N_\beta=2$ with $r=1$. Although the convergence and fluctuations around the theoretical expression given in Eq.~(\ref{eq:Sirr_r}) (solid horizontal lines) depend on the specific values $r$ and $N_\beta$, the theoretical expression correctly captures the saturation value. The behavior of the irreversible entropy is equivalent to the one shown in Fig.~\ref{fig:transient}(a) since $\langle S_{\rm irr}\rangle=\beta\langle W_{\rm irr}\rangle$.  
 
Finally, in Fig.~\ref{fig:kappa3} we show the behavior of the third cumulant $\kappa_3^{1/3}$ of the work probability distribution $P(W)$ as a function of the number of thermal excitations in the initial state $N_\beta$ and the protocol exponent $r$. The third cumulant grows as both $N_\beta$ and $r$ increase, which complements the results shown in Fig.~\ref{fig2}(a). We also note that the first and second cumulants $\kappa_1$ and $\kappa_2$ follow a very similar behavior (see caption of Fig.~\ref{fig:kappa3}).

\section{Irreversible work in a transverse-field Ising model}\label{app:Ising}

 \begin{figure}[t!]
 \centering
 \includegraphics[width=0.85\linewidth,angle=-0]{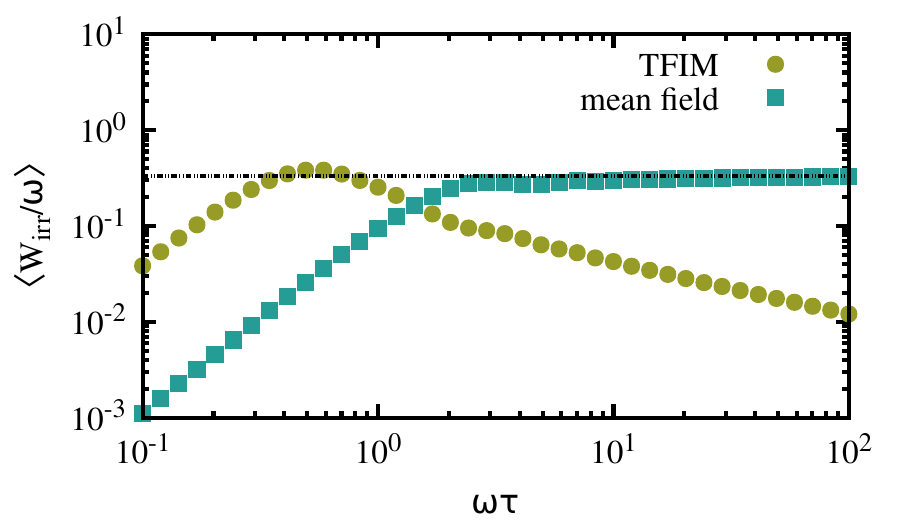}
 \caption{\label{fig:app}\small{Average irreversible work $\langle W_{\rm irr}\rangle$ for the TFIM with $N=200$ spins and for the mean-field system in the exact thermodynamic limit, considered in the main text, cf. Eq.~(\ref{eq:H0}), as a function of the driving time $\tau$ and for a linear ramp $r=1$. Both systems are initialized in their corresponding ground states at $g(0)$. The cycle from $g(0)=0$ to $g(\tau)=1$ and back to $g(2\tau)=0$ is performed in a total time $2\tau$. The scaling of $\langle W_{\rm irr}\rangle$ for the TFIM follows the Kibble-Zurek scaling, $\langle W_{\rm irr}\rangle \propto \tau^{-d\nu r/(1+z\nu r)}$ with $d=\nu=z=1$, which in this case leads to $\langle W_{\rm irr}\rangle\propto \tau^{-1/2}$, while the mean-field model saturates to $\langle W_{\rm irr}\rangle=\omega \sinh^2[|s|]$ even in the limit $\tau\rightarrow\infty$ (dashed line), as explained in the main text.}}
 \end{figure}

 As mentioned in the main text, non-zero dimensional critical systems feature a different behavior of the average irreversible work $\langle W_{\rm irr}\rangle$ upon the completion of a finite-time cycle reaching the critical point. For slow cycles, the irreversible work acquires a Kibble-Zurek scaling, $\langle W_{\rm irr}\rangle\propto \tau^{-d\nu r/(1+z\nu r)}$, where $d$ is the dimensionality of the system. For mean-field critical systems, $d=0$ and thus the $\langle W_{\rm irr}\rangle$ saturates to a constant value, as explained in the main text. Here, we compare the average irreversible work of the paradigmatic transverse field one-dimensional quantum Ising model (TFIM) with the results obtained for the mean-field model, cf.~Eq.~(\ref{eq:H0}). The TFIM describes a chain of $N$ spins with a transverse field strength $g$, whose Hamiltonian can be written as
  \begin{equation}
  \hat{H}(t)=-J\sum_{n=1}^{N} \hat{\sigma}_n^z\hat{\sigma}_{n+1}^z - g\pap{t}\sum_{n=1}^{N}\hat{\sigma}_n^{x},
  \end{equation}
 where $J>0$ is the spin-spin hopping interaction. For simplicity, we set $J=\omega=1$ and we consider periodic boundary conditions, $\hat{\sigma}^\alpha_{N+1}=\hat{\sigma}^\alpha_1$ with $\alpha=x,y,z$ that refer to the spin-$1/2$ Pauli matrices. The TFIM exhibits a second-order quantum phase transition at $g_c = J$ between a paramagnetic phase ($g>J$) and a ferromagnetic phase ($g<J$), where the energy gap between ground and first excited states vanishes in the thermodynamic limit, $N\rightarrow \infty$. Note that at the critical point of a mean-field quantum phase transition, the energy gap between all the eigenstates vanish, which may anticipate a different dynamical behavior.  The TFIM can be diagonalized employing the standard Jordan-Wigner transformation, which allows us to solve exactly the dynamics as a set of coupled Landau-Zener systems (see, for example, Refs.~\cite{Dziarmaga:05, Zurek:05}). We consider a similar time-dependent magnetic field function given by Eq.~(\ref{eq:gt}), i.e.~from $g(0)=0$ to $g(\tau)=g_c=1$, and back to the initial magnetic field $g(2\tau)=0$. The resulting average irreversible work $\langle W_{\rm irr}\rangle$ is plotted in Fig.~(\ref{fig:app}) for $N=200$ spins starting in its ground state at $g(0)=0$, together with $\langle W_{\rm irr}\rangle$ for the mean-field model in Eq.~(\ref{eq:H0}), employing a linear ramp ($r=1$). Similar results can be found for other choices of $r$ and system sizes $N$ for the TFIM (see also~\ref{app:transient} for the saturation of $\langle W_{\rm irr}\rangle$ for other values of $r$ and thermal occupation in the mean-field system). Note that the scaling behavior of $\langle W_{\rm irr}\rangle$ for the TFIM decreases for slower cycles, namely when $\tau$ becomes large. Finally, let us mention that the saturation of $\langle W_{\rm irr}\rangle$ for slow cycles is in agreement with the results reported in~\cite{Defenu:18, Gherardini:24}.

  \section{Squeezed thermal state}\label{app:ST}
 The cycle considered in this paper squeezes the initial state. Thus, for an initial thermal state $\hat{\rho}_\beta$ at inverse temperature $\beta$, the final state is $\hat{\rho}(2\tau)=\hat{\mathcal{S}}(s)\hat{\rho}_\beta\hat{\mathcal{S}}^\dagger(s)$ being $\hat{\mathcal{S}}(s)$ the squeezing operator with parameter $s=|s|e^{i\phi_s}$. This form allows us to explicitly write the work probability distribution as
  \begin{equation}
  P(2k\omega) = \sum_{n\geq 0}\sum_{m=2k+n,\,m\geq 0} \frac{N_\beta^{n}(\omega)}{(1+N_\beta(\omega))^{n+1}}S_{n,m}(|s|), 
  \end{equation}
  where $N_\beta(\omega)=(e^{\beta\omega}-1)^{-1}$ is the amount of thermal excitations, and $S_{n,m}(|s|)$ is the photon number distribution of a squeezed number state $\ket{n}$. $S_{n,m}(|s|)$ reads as~\cite{Kim:89}
 \begin{eqnarray}
   S_{n,m}(|s|) &= \left| \bra{m}\hat{\mathcal{S}}(s)\ket{n}\right|^2, \nonumber\\
  &=\frac{n!m!}{2^{n-m}}\frac{\tanh^{n-m}[|s|]}{\cosh^{2m+1}[|s|]}\big| Q(|s|,m,n) \big|^2,
 \end{eqnarray}
  for $n-m$ even, while $S_{n,m}(|s|)=0$ otherwise. The function $Q(|s|,m,n)$ is given by
  \begin{equation}
  Q(|s|,m,n)=\sum_{k=(m-n)/2}^{m/2}\frac{(-1)^k\sinh^{2k}[|s|]}{4^{k}k!(m-2k)!(k+(n-m)/2)!}.
  \end{equation}
  From this expression, we can find an analytical and universal work probability distribution upon particularizing $|s|$ as given in Eq.~(\ref{eq:s}) for sufficiently slow cycles, $2\omega\tau\gtrsim 10$. However, since the expressions are intricate for a general initial thermal state, we provide closed expressions for the simple case of an initial vacuum state.

  Then, the photon number distribution of the squeezed thermal state $\hat{\rho}(2\tau)$ is 
 \begin{equation}
  p_n(\beta,s)=\sum_{m\geq 0} \frac{N_\beta^n}{(1+N_\beta)^{n+1}}S_{m,n}(|s|),
  \end{equation}
  so that $\sum_np_n(\beta,s)=1$. This expression is required to find the quantum coherence contribution $C$ to the irreversible entropy, as explained in Sec.~\ref{s:coh}. Notice that an equivalent expression for the photon number distribution $p_n(\beta,s)$ can also be found in~\cite{Yimin:97}:
  \begin{eqnarray}
  p_n(\beta,s)= \frac{2\ n!\ell^{2n}}{h^{1/2}}\left\{
 \begin{array}{cc}
 \displaystyle{\sum_{k=0}^{n/2}\frac{q^{2k}h^{-4k}}{(2k)!\left[(n/2-k)! \right]^{2}}} &{\rm for }\;  n \ {\rm even}\\
 \displaystyle{ \sum_{k=0}^{(n-1)/2}\frac{q^{2k+1}h^{-4k-2}}{(2k+1)!\left[((n-1)/2-k)! \right]^{2}} } &{\rm for } n\  {\rm odd}
 \end{array}
 \right.
  \end{eqnarray}
  with $\mu=\frac{1+e^{-\beta\omega}}{1-e^{-\beta\omega}}$, $h=(1+\mu e^{2|s|})(1+\mu e^{-2|s|})$, $\ell=\sqrt{\mu \sinh[2|s|]/h}$ and $q=(\mu^2-1) h^{-1}$. 

\section*{References}
\bibliographystyle{iopart-num}
\bibliography{paper.bib}
\end{document}